\documentclass[aps,prb]{revtex4}
\usepackage{graphicx}

\begin{document}
\pagestyle{empty}
\title{ Theory  of the interaction forces and the heat transfer  between moving bodies mediated
by the fluctuating electromagnetic field }

\author{A.I. Volokitin$^{1,2}$\footnote{Corresponding author.
\textit{E-mail address}:alevolokitin@yandex.ru}    and B.N.J.
Persson$^1$} \affiliation{$^1$Institut f\"ur
Festk\"orperforschung, Forschungszentrum J\"ulich, D-52425,
Germany} \affiliation{$^2$Povolzhskaya State Academy of
Telecommunication and Informatics, 443010 Samara, Russia}

\begin{abstract}
Within the framework of unified approach we study the
Casimir-Lifshitz interaction,  the van der Waals friction force
and the radiative heat transfer at nonequilibrium conditions, when
the interacting bodies are at different temperatures, and they
move relative to each other with the arbitrary velocity $V$. The
analysis is focused on the surface-surface and surface-particle
configuration. Our approach is based on the exact solution of
electromagnetic problem about the determination of the fluctuating
electromagnetic field in the vacuum gap between two flat parallel
surfaces moving relative to each other with the arbitrary velocity
$V$. The velocity dependence of the considered phenomena is
determined by Doppler shift and can be strong for resonant photon
tunneling between surface modes.
  We show that
relativistic effects give rise to a mixing of the contributions
from the electromagnetic waves with different polarization to the
heat transfer and the interaction forces. We find that these
effects are of the order $(V/c)^2$. The limiting case when one of
the bodies is sufficiently rarefied gives the heat transfer and
the interaction forces between a moving small particle and a
surface.
 We also calculate
the friction force acting on a particle moving with an arbitrary velocity relative to the black body radiation.

\end{abstract}

\maketitle

\section{Introduction}

All bodies are surrounded by a fluctuating electromagnetic field due to the thermal and quantum fluctuations
of the charge  and  current density inside
the bodies. Outside the bodies this fluctuating electromagnetic field exists partly in the form of propagating
electromagnetic waves and partly in
the form of evanescent  waves. The theory of the fluctuating electromagnetic field was developed by Rytov
\cite{Rytov53,Rytov67,Rytov89}.
A great variety of phenomena such as Casimir-Lifshitz forces \cite{Lifshitz54},  near-field radiative heat
transfer \cite{Polder}, and friction forces
\cite{Volokitin99}
can be described using this theory.

Lifshitz \cite{Lifshitz54} used the Rytov's theory to formulate a
very general theory of the dispersion interaction in the framework
of the statistical physics and macroscopic electrodynamics. The
Lifshitz theory  provides a common tool to deal with dispersive
forces in different field of science (physics, biology, chemistry)
and technology.

The Lifshitz theory is formulated for systems at thermal
equilibrium. At present there is an interest in the study of
systems out of the thermal equilibrium (see \cite{Pitaevskii08}
and reference therein). The principal interest in the study of
systems out of thermal equilibrium is connected to the possibility
of tuning the interaction in both strength and sign
\cite{Antezza05,Antezza06}. Such systems also give a way to
explore the role of thermal fluctuations usually masked at thermal
equilibrium by the $T=0$ K component which dominates the
interaction up to very large distances, where the actual total
force result in to be very small. The Casimir-Lifshitz force was
measured at very large distances and it was shown that the thermal
effects of the Casimir-Lifshitz interaction are in agreement with
the theoretical prediction \cite{Antezza05}. This measurement was
done out of thermal equilibrium, where thermal effects are
stronger.

  Further thermal
non-equilibrium
 effects were explored by Polder and Van Hove \cite{Polder}, who calculated the heat-flux between two
  parallel plates.
 At present there is
an increasing  interest for studying near-field radiative heat
transfer
\cite{Pendry99,Volokitin01a,Volokitin04,Volokitin03,Mulet01,Mulet02}
which is connected with the development of a near-field scanning
thermal microscope \cite{Kittel}. The existing studies are limited
mostly by the case when the interacting bodies  have different
temperatures but they are  at  rest.
 For recent review of near-field radiative heat transfer between bodies, which are at rest, see
\cite{Joulain05,RMP07}.

Another non-equilibrium effects are realized for bodies moving
relative to each other. In \cite{Volokitin99} we used the
dynamical modification of the Lifshitz theory to calculate the
friction force between two plane parallel surfaces in  relative
motion with velocity $V$. The calculation of the van der Waals
friction is more complicated than of the Casimir-Lifshitz force
and the radiative heat transfer because it requires the
determination of the electromagnetic field between moving
boundaries. The solution can be found by writing the boundary
conditions on the surface of each body in the rest reference frame
of this body. The relation between the electromagnetic fields in
the different reference frames is determined by the Lorenz
transformation.    In \cite{Volokitin99} the electromagnetic field
in the vacuum gap between the bodies was calculated to linear
order in  $V/c$. It was shown that linear terms in the
electromagnetic field give the contribution to the friction force
of the order $(V/c)^2$. Thus, these linear terms were neglected in
\cite{Volokitin99} and the resulting  formula for friction force
is accurate to order $(V/c)^2$. The same approximation was used in
\cite{Volokitin01b} to calculate the frictional drag between
quantum wells, and in \cite{Volokitin03a,Volokitin03b} to
calculate the friction force between plane parallel surfaces in
normal relative motion.  For a recent review of the van der Waals
friction see \cite{RMP07}.

In this paper  within the framework of unified approach we study
the Casimir-Lifshitz interaction,  the van der Waals friction
force and the radiative heat transfer at nonequilibrium
conditions, when the interacting bodies are at different
temperatures, and they move relative to each other with the
arbitrary velocity $V$. Our study is focused on the
surface-surface and surface-particle configuration. In comparison
with previous studies we consider more general nonequilibrium
conditions. In the existing literature the Casimir-Lifshitz
interaction and the radiative heat transfer for the
surface-surface configuration were studied only for the systems
out of the thermal equilibrium
\cite{Pitaevskii08,Joulain05,RMP07}. The van der Waals friction is
studied for this configuration only for systems at thermal
equilibrium \cite{RMP07}. In Sec. \ref{theory} we calculate the
fluctuating electromagnetic field in the vacuum gap between two
plane parallel surfaces,
 moving in parallel relative to each other with arbitrary velocity $V$.
 In comparison with previous calculations
 \cite{Volokitin99,Volokitin01b,Volokitin03a,
Volokitin03b} we do not make any approximation in the Lorentz
transformation of the electromagnetic field by means of which we
can determine the field in one inertial reference frame, knowing
the same field in another reference frame. Thus, our solution of
the electromagnetic problem is exact. Knowing the electromagnetic
field we calculate the stress tensor and the Poynting vector which
determine the interaction force and the heat transfer,
respectively. We calculate the friction force and the conservative
force, and the radiative heat transfer  in Sects. \ref{friction},
\ref{force} and \ref{heat}, respectively. Upon going to the limit
when one of the bodies is rarefied we  obtain the interaction
force and the heat transfer for a small particle-surface
configuration.   In  Sec. \ref{bb}  we calculate the friction
force on a small particle moving relative to black body radiation.
The same problem was considered in \cite{Mkrtchian}. In comparison
with this study  our treatment is relativistic and we take into
account the contribution not only from the electric dipole moment
but also from the magnetic moment of the particle. Recently
\cite{Chapius08} it was shown that the magnetic moment gives the
most important contribution to the near-field radiative heat
transfer for metallic particles. The same is true for the friction
force.    The conclusions are given in Sec. \ref{conclusions}.

\section{Calculation of the fluctuating electromagnetic field \label{theory}
}

\begin{figure}
\includegraphics[width=0.45\textwidth]{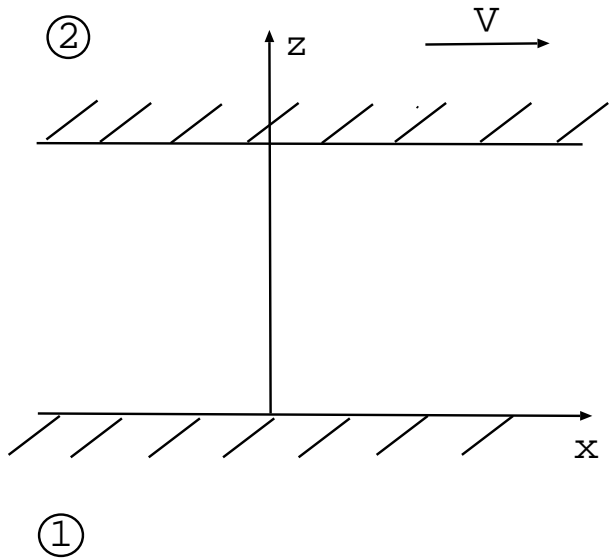}
\caption{\label{Fig1}
Two semi-infinite bodies with plane parallel surfaces separated by a
distance $d$.
The upper solids moves parallel to  other with velocity $V$.
}
\end{figure}

We consider two semi-infinite solids having flat parallel surfaces separated
by a distance $d$ and moving with velocity $V$ relative to each other, see
Fig. \ref{Fig1}.
We introduce the two coordinate systems $K$ and $K^{\prime }$ with
coordinate axes $xyz$ and $x^{\prime }y^{\prime }z^{\prime }$. In
the $K$ system body \textbf{1} is at rest while body \textbf{2} is
moving with the velocity $V$ along the $x-$ axis ($xy$ and
$x^{\prime }y^{\prime }$ planes are in the surface of body
\textbf{1, }$x$ and $ x^{\prime}$- axes have the same direction,
and the $z$ and $z^{\prime}-$ axes point toward body \textbf{2}).
In the $K^{\prime}$ system body \textbf{2} is at rest while body
\textbf{1} is moving with velocity $-V$ along the $x-$ axis. Since
the system is translational invariant in the $\mathbf{x}=(x,y)$
plane, the electromagnetic field can be represented by the Fourier
integrals
\begin{eqnarray}
\mathbf{E}(\mathbf{x},z,t)=\int_{-\infty}^{\infty} d\omega\int \frac{d^2q}{(2\pi)^2}e^{i\mathbf{q}
\cdot \mathbf{x}-i\omega t}
\mathbf{E}(\mathbf{q},\omega,z),   \\
\mathbf{B}(\mathbf{x},z,t)=\int_{-\infty}^{\infty} d\omega\int \frac{d^2q}{(2\pi)^2}e^{i\mathbf{q}
\cdot \mathbf{x}-i\omega t}
\mathbf{B}(\mathbf{q},\omega,z),
\end{eqnarray}
where $\mathbf{E}$ and $\mathbf{B}$ are the electric and magnetic induction field, respectively,
and $\mathbf{q}$ is the two-dimensional wave vector in $xy$- plane. After
Fourier transformation it is convenient to decompose the electromagnetic field into  $s$- and $p$- polarized components.
For the $p$- and $s$ -polarized electromagnetic waves the electric field $\mathbf{E}(\mathbf{q},\omega,z)$ is in plane of incidence,
and perpendicular to that plane, respectively. In the vacuum
gap between the bodies  the electric field $\mathbf{E}(\mathbf{q},\omega,z)$,  and the  magnetic induction field
$\mathbf{B}(\mathbf{q},\omega,z)$ can be written in the form
\begin{equation}
\mathbf{E}(\mathbf{q},\omega ,z)= \left( v_s\hat{n}_s +
v_p\hat{n}_p^+\right)e^{ik_zz} +\left( w_s\hat{n}_s +
w_p\hat{n}_p^-\right)e^{-ik_zz}
  \label{one}
\end{equation}
\begin{eqnarray}
\mathbf{B}(\mathbf{q},\omega ,z)=\frac {c}{\omega}\left[ \left([\mathbf{k}^{+}
\times \hat{n}_s]v_s + [ \mathbf{k}^{+}\times\hat{n}_p^{+}]v_p\right)e^{ik_zz}  \right.     \nonumber   \\
\left. + \left( [ \mathbf{k}^{-}\times \hat{n}_s]w_s + [
\mathbf{k}^{-}\times \hat{n}_p^{-}]w_p\right)e^{-ik_zz} \right]
 \label{two}
\end{eqnarray}
where $\mathbf{k}^{\pm}=\mathbf{q}\pm \hat{z}k_z,\,
k_z=((\omega/c)^2-q^2)^{1/2},\,\hat{n}_s=[\hat{z}\times\hat{q}] = (-q_y, q_x,0)/q,\,\hat{n}_p^{\pm}=[\hat{k}^{\pm}\times
\hat{n}_s]=(\mp q_xk_z,\mp q_yk_z,q^2)/(kq),\, k=\omega/c $.
At the surfaces of the bodies the amplitude
of the outgoing electromagnetic wave must be equal to the amplitude of the reflected
wave plus the amplitude of the radiated wave. Thus, the boundary conditions
for the electromagnetic
field at $z=0$ in the $K$- reference frame
can be written in the form
\begin{equation}
v_{p(s)}=R_{1p(s)}(\omega,q )w_{p(s)}+E^{f}_{1p(s)}(\omega,q)
\label{three}
\end{equation}
where $R_{1p(s)}(\omega )$ is the reflection amplitude for surface
\textbf{1} for the $p(s)$ - polarized electromagnetic field, and
where $ E^{f}_{1p(s)}(\omega )$ is the amplitude of the
 fluctuating electric field radiated  by body \textbf{1} for a $p(s)$-polarized wave.
In the $K^{^{\prime }}$- reference frame  the electric field can be written in the form
\begin{equation}
\mathbf{E}^{\prime}(\mathbf{q}^{\prime},\omega^{\prime} ,z)=
\left( v_s^{\prime}\hat{n}_s^{\prime} +
v_p^{\prime}\hat{n}_p^{\prime +}\right)e^{ik_zz} +\left(
w_s^{\prime}\hat{n}_s^{\prime} + w_p^{\prime}\hat{n}_p^{\prime
-}\right)e^{-ik_zz}
  \label{four}
\end{equation}
where $\mathbf{q}^{\prime}=(q_x^{\prime},q_y,0),\,q_x^{\prime}=(q_x - \beta k)\gamma,\,\omega^{\prime}=
(\omega - Vq_x)\gamma,\,
\gamma=1/\sqrt{1-\beta^2},\,\beta = V/c,\, \hat{n}^{\prime}_s =
(-q_y, q^{\prime}_x,0)/q^{\prime},\,\hat{n}^{\prime \pm}_{p}=
(\mp q^{\prime}_xk_z,\mp q_yk_z,q^{\prime 2})/(k^{\prime}q^{\prime}),$
\[
q^{\prime}=\gamma\sqrt{q^2-2\beta kq_x+\beta^2(k^2-q_y^2)}.
\]
The boundary conditions at $z=d$ in the $K^{^{\prime }}$- reference frame can be written in a form similar to
Eq. (\ref{three}):
\begin{equation}
w^{\prime}_{p(s)}=e^{2ik_zd}R_{2p(s)}(\omega^{\prime},q^{\prime} )v^{\prime}_{p(s)}
+e^{ik_zd}E^{\prime f}_{2p(s)}(\omega^{\prime},q^{\prime} ),
\label{five}
\end{equation}
where $R_{2p(s)}(\omega )$ is the reflection amplitude for surface
\textbf{2} for $p(s)$ - polarized electromagnetic field, and where
$ E^{f}_{2p(s)}(\omega )$ is the amplitude of the
 fluctuating electric field radiated by  body \textbf{2} for a $p(s)$-polarized wave.
A Lorentz transformation for the electric field gives
\begin{equation}
E_x^{\prime}=E_x,\,E_y^{\prime} = (E_y-\beta B_z)\gamma,\, E_z^{\prime} = (E_z+\beta B_y)\gamma
\label{six}
\end{equation}
Using Eqs. (\ref{one},\ref{two},\ref{four}) and (\ref{six}) we get
\begin{equation}
v_p^{\prime} = \frac{k^{\prime}\gamma}{kqq^{\prime}}\left[-\beta k_zq_yv_s + (q^2 - \beta kq_x)v_p\right],
\label{seven}
\end{equation}
\begin{equation}
w_p^{\prime} = \frac{k^{\prime}\gamma}{kqq^{\prime}}\left[\beta k_zq_yw_s + (q^2 - \beta kq_x)w_p\right],
\label{eight}
\end{equation}
\begin{equation}
v_s^{\prime} = \frac{k^{\prime}\gamma}{kqq^{\prime}}\left[\beta k_zq_yv_p + (q^2 - \beta kq_x)v_s\right],
\label{nine}
\end{equation}
\begin{equation}
w_s^{\prime} = \frac{k^{\prime}\gamma}{kqq^{\prime}}\left[-\beta k_zq_yw_p + (q^2 - \beta kq_x)w_s\right].
\label{ten}
\end{equation}
Substituting Eqs. (\ref{seven}-\ref{ten}) in Eq. (\ref{five}) and using Eq. (\ref{three}) we get
\[
(q^2 - \beta kq_x)D_{pp}w_p+\beta k_zq_yD_{sp}w_s
\]
\begin{equation}
=e^{2ik_zd}R_{2p}^{\prime}\left[(q^2 - \beta kq_x)E_{1p}^f-\beta k_zq_yE_{1s}^f\right] +
\frac{kqq^{\prime}}{k^{\prime}\gamma}e^{ik_zd}E_{2p}^{\prime f},
\label{eleven}
\end{equation}
\[
(q^2 - \beta kq_x)D_{ss}w_s - \beta k_zq_yD_{ps}w_p
\]
\begin{equation}
=e^{2ik_zd}R_{2s}^{\prime}\left[(q^2 - \beta kq_x)E_{1s}^f + \beta k_zq_yE_{1p}^f\right] +
\frac{kqq^{\prime}}{k^{\prime}\gamma}
e^{ik_zd} E_{2s}^{\prime f},
\label{twelve}
\end{equation}
where
\[
D_{pp} = 1 - e^{2ik_zd}R_{1p}R_{2p}^{\prime},\, D_{ss} = 1 - e^{2ik_zd}R_{1s}R_{2s}^{\prime},
\]
\[
D_{sp} = 1 + e^{2ik_zd}R_{1s}R_{2p}^{\prime},\, D_{ps} = 1 + e^{2ik_zd}R_{1p}R_{2s}^{\prime},
\]
$R_{2p(s)}^{\prime}=R_{2p(s)}(\omega^{\prime},q^{\prime})$.
From Eqs. (\ref{eleven},\ref{twelve}) and (\ref{three}) we get
\[
w_p = \Big\{\left[(q^2 - \beta kq_x)^2R_{2p}^{\prime}D_{ss} - \beta^2 k_z^2q_y^2R_{2s}^{\prime}D_{sp}\right]
E_{1p}^fe^{2ik_zd}
\]
\[
-\beta k_zq_y(q^2 - \beta kq_x)(R_{2p}^{\prime} +
R_{2s}^{\prime})E_{1s}^fe^{2ik_zd}
\]
\begin{equation}
+ \frac{kqq^{\prime}}{k^{\prime}\gamma}\left[(q^2 - \beta
kq_x)D_{ss}E_{2p}^{\prime f}- \beta k_zq_yD_{sp}E_{2s}^{\prime
f}\right]e^{ik_zd} \Big\}\Delta^{-1}, \label{thirteen}
\end{equation}
\[
v_p = \Big\{\left[(q^2 - \beta kq_x)^2D_{ss} + \beta^2 k_z^2q_y^2D_{sp}\right]
E_{1p}^f
\]
\[
-\beta k_zq_y(q^2 - \beta kq_x)R_{1p}(R_{2p}^{\prime} + R_{2s}^{\prime})e^{2ik_zd}E_{1s}^f
\]
\begin{equation}
+ \frac{kqq^{\prime}}{k^{\prime}\gamma}R_{1p}\left[(q^2 - \beta kq_x)D_{ss}E_{2p}^{\prime f}-
\beta k_zq_yD_{sp}E_{2s}^{\prime f}\right]e^{ik_zd}
\Big\}\Delta^{-1},
\label{fourteen}
\end{equation}

\[
w_s = \Big\{\left[(q^2 - \beta kq_x)^2R_{2s}^{\prime}D_{pp} - \beta^2 k_z^2q_y^2R_{2p}^{\prime}D_{ps}\right]
E_{1s}^fe^{2ik_zd}
\]
\[
+\beta k_zq_y(q^2 - \beta kq_x)(R_{2p}^{\prime} + R_{2s}^{\prime})E_{1p}^fe^{2ik_zd}
\]
\begin{equation}
+ \frac{kqq^{\prime}}{k^{\prime}\gamma}\left[(q^2 - \beta kq_x)D_{pp}E_{2s}^{\prime f}+
\beta k_zq_yD_{ps}E_{2p}^{\prime f}\right]e^{ik_zd}
\Big\}\Delta^{-1},
\label{fifteen}
\end{equation}
\[
v_s = \Big\{\left[(q^2 - \beta kq_x)^2D_{pp} + \beta^2 k_z^2q_y^2D_{ps}\right]
E_{1s}^f
\]
\[
+\beta k_zq_y(q^2 - \beta kq_x)R_{1p}(R_{2p}^{\prime} + R_{2s}^{\prime})e^{2ik_zd}E_{1p}^f
\]
\begin{equation}
+ \frac{kqq^{\prime}}{k^{\prime}\gamma}R_{1s}\left[(q^2 - \beta
kq_x)D_{pp}E_{2s}^{\prime f}+ \beta k_zq_yD_{ps}E_{2p}^{\prime
f}\right]e^{ik_zd} \Big\}\Delta^{-1}, \label{sixteen}
\end{equation}
where
\[
\Delta = (q^2 - \beta kq_x)^2D_{ss}D_{pp} + \beta^2k_z^2q_y^2D_{ps}D_{sp}.
\]

The fundamental characteristic
of the fluctuating electromagnetic field is the correlation function, determining
the average product of amplitudes $E^{f}_{p(s)}(\mathbf{q},\omega)$.
According to the
general theory of the fluctuating electromagnetic field
(see for a example \cite{RMP07}):
\[
<|E^{f}_{p(s)}(\mathbf{q},\omega)|^2>=\frac{\hbar \omega^2}
{2c^2|k_z|^2} \left(n(\omega)+\frac{1}{2}\right)
[(k_z+k_z^*)(1-|R_{p(s)}|^2)
\]
\begin{equation}
+(k_z-k_z^*)(R_{p(s)}^*-R_{p(s)})]  \label{seventeen}
\end{equation}
where $<...>$ denote statistical average over the random field.
We note that $k_z$ is  real for $q<\omega/c$ (propagating waves), and purely
imaginary for  $q>\omega/c$ (evanescent waves). The the Bose-Einstein factor
\[
n(\omega )=\frac 1{e^{\hbar \omega /k_BT}-1}.
\]
Thus for $q<\omega/c$
and $q>\omega/c$ the
correlation functions are determined by the first and the second
terms in Eq. (\ref{seventeen}), respectively.

\section{Calculation of the  friction force \label{friction}}

The force which acts  on the surface of body \textbf{1} can be calculated from the Maxwell stress tensor
$\sigma_{ij}$, evaluated at $z=0$:
\[
\sigma _{ij} =\frac 1{4\pi }\int_0^\infty d\omega \int \frac{d^2q}{(2\pi)^2}
\Big[ <E_iE_j^*> + <E_i^*E_j> + <B_iB_j^*> + <B_i^*B_j>
\]
\begin{equation}
 - \delta_{ij}(<\mathbf{E\cdot E^*}> + <\mathbf{B\cdot B^*}>)\Big] _{z=0}   \label{3one}
\end{equation}
Using Eqs. (\ref{one},\ref{two}) for the $x$ - component of the force we get
\[
\sigma_{xz} =\frac 1{4\pi }\int_0^{\infty }d\omega \int \frac{d^2q}{(2\pi)^2}\frac{q_x}{k^2} \left[
(k_z+k_z^{*})\left (\left\langle \mid w_p\mid
^2\right\rangle +\left\langle \mid w_s\mid
^2\right\rangle
\right. \right.
\]
\begin{equation}
\left.\left. -\left\langle \mid v_p\mid ^2\right\rangle -
\left\langle \mid v_s\mid ^2\right\rangle \right)
+(k_z-k_z^{*})\left\langle w_pv_p^* + w_sv_s^*
 - c.c\right\rangle \right]
\label{2two}
\end{equation}
Substituting Eqs.  (\ref{thirteen}-\ref{sixteen}) for the amplitudes of the
electromagnetic field in Eq. (\ref{2two}),  and performing averaging over the fluctuating electromagnetic field with the help of
Eq. (\ref{seventeen}), we get the $x$-component of the force
\[
F_x = \sigma_{xz} =\frac \hbar {8\pi ^3}\int_0^\infty d\omega \int_{q<\omega /c}d^2q\frac{q_x}{|\Delta|^2}[(q^2 - \beta kq_x)^2 + \beta^2k_z^2q_y^2]
\]
\[
\times  [(q^2 - \beta kq_x)^2(1-\mid R_{1p}\mid ^2)(1-\mid R_{2p}^{\prime }\mid ^2)|D_{ss}|^2
\]
\[
+\beta^2k_z^2q_y^2(1-\mid R_{1p}\mid ^2)(1-\mid R_{2s}^{\prime }\mid ^2)|D_{sp}|^2 + (p\leftrightarrow s)]
\left( n_2(\omega^{\prime})-n_1(\omega )\right)
\]
\[
+\frac \hbar {2\pi ^3}\int_0^\infty d\omega \int_{q>\omega
/c}d^2q\frac{q_x}{|\Delta|^2}[(q^2 - \beta kq_x)^2 + \beta^2k_z^2q_y^2]
e^{-2\mid k_z\mid d}
\]
\[
\times [(q^2 - \beta kq_x)^2 \mathrm{Im}R_{1p}\mathrm{Im}R_{2p}^{\prime}|D_{ss}|^2- \beta^2k_z^2q_y^2
\mathrm{Im}R_{1p}\mathrm{Im}
R_{2s}^{\prime}|D_{sp}|^2
\]
\begin{equation}
+ (p\leftrightarrow s)]\left( n_2(\omega^{\prime})-n_1(\omega
)\right),   \label{2three}
\end{equation}
where
\[
n_i(\omega )=\frac 1{e^{\hbar \omega /k_BT_i}-1},
\]
where $T_1$ and $T_2$ are the temperatures for bodies \textbf{1}
and \textbf{2}, respectively. The symbol $(p\leftrightarrow s)$
denotes the terms which can be obtained from the preceding terms
by permutation of the indexes $p$ and $s$. The first term in Eq.
(\ref{2three}) represents the contribution to the friction from
propagating waves ($q<\omega /c$), and the second term  from the
evanescent waves ($q>\omega /c$). If in Eq. (\ref{2three})  one
neglects the terms of the order $\beta^2$ then the contributions
from waves with $p$- and $s$- polarization will be separated. In
this case Eq. (\ref{2three}) is reduced to the formula obtained in
\cite{Volokitin99}. Thus, to the order $\beta^2$ the mixing of
waves with different polarization can be neglected, what agrees
with the results obtained in \cite{Volokitin99}. At $T=0$ K the
propagating waves do not contribute to friction but the
contribution from evanescent waves is not equal to zero. Taking
into account that $n(-\omega)= -1 - n(\omega)$ from Eq.
(\ref{2three}) we get friction mediated by the evanescent
electromagnetic waves at zero temperature (in literature this type
of friction is named as quantum friction \cite{Pendry97})
\[
F_x =
- \frac \hbar {\pi ^3}\int_0^\infty dq_y \int_0^\infty dq_x \int_0^{q_xV}
d\omega \frac{q_x}{|\Delta|^2}[(q^2 - \beta kq_x)^2 + \beta^2k_z^2q_y^2]
e^{-2\mid k_z\mid d}
\]
\begin{equation}
\times [(q^2 - \beta kq_x)^2 \mathrm{Im}R_{1p}\mathrm{Im}R_{2p}^{\prime}|D_{ss}|^2- \beta^2k_z^2q_y^2
\mathrm{Im}R_{1p}\mathrm{Im}
R_{2s}^{\prime}|D_{sp}|^2 + (p\leftrightarrow s)]
\label{zerotemfr}
\end{equation}
Fig. 2 shows the dependence of the frictional stress between
semi-infinite bodies on the velocity $V$ at different separation
$d$. In the calculations the Fresnel formulas for the reflection
amplitude were used with the Drude permittivity $\varepsilon$ for
copper. The frictional stress initially increases with velocity,
reaches a maximum at $V\sim d k_BT/\hbar$, and then decreases at
large values of the velocity.
\begin{figure}
\includegraphics[width=0.45\textwidth]{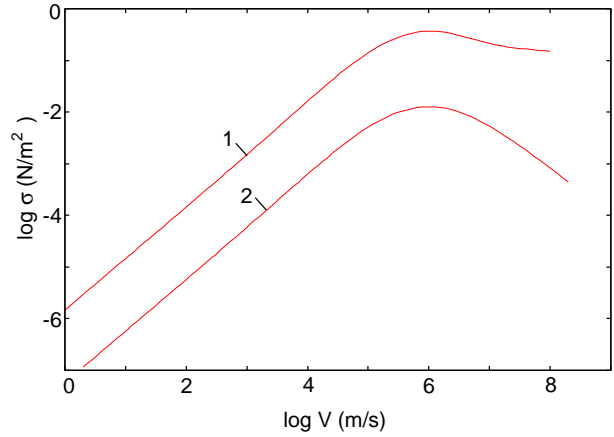}
\caption{\label{Fig2}
The velocity dependence of the frictional stress acting between two semi-infinite bodies  at
$d=10$nm (1) and $d=100$nm (2),
with parameters chosen to correspond to copper ($\tau
^{-1}=2.5\cdot 10^{13}$ s$^{-1}$, $\omega _p=1.6\cdot 10^{16}$
s$^{-1}$).  (The base of the logarithm is 10)
}

\end{figure}

The  friction force acting on a small particle moving in  parallel
to a flat surface can be obtained from the friction between two
semi-infinite bodies in the limit when one of the bodies is
sufficiently rarefied. For $d\ll \lambda_T=c\hbar/k_BT$,  in Eq.
(\ref{2three}) we can neglect by the first term and in the second
term we can integrate over the whole $q$-plane, and put
$k_z\approx iq$. We will assume that the rarefied body consists of
small metal particles which  have  the electric dipole moment and
the magnetic moment. The dielectric permittivity and magnetic
permeability of this body, say body \textbf{2}, is close to the
unity, i.e. $\varepsilon _2-1\rightarrow 4\pi n\alpha_E\ll 1$ and
$\mu _2-1\rightarrow 4\pi n\alpha_H\ll 1$, where $n$ is the
concentration of particles in body \textbf{2},  $\alpha _E$ and
$\alpha _H$ are  their electric and magnetic susceptibilities. To
linear order in the concentration $n$ the reflection amplitudes
are
\[
R_{2p} = \frac{\varepsilon _2k_z - \sqrt{\varepsilon _2\mu_2 k^2 - q^2}}{\varepsilon _2k_z +
\sqrt{\varepsilon _2\mu_2 k^2 - q^2}}
\approx \frac{\varepsilon _2-1}{\varepsilon _2+1}\approx 2\pi n\alpha_E
\]
\[
R_{2s} = \frac{\mu _2k_z - \sqrt{\varepsilon _2\mu_2 k^2 - q^2}}{\mu _2k_z +
\sqrt{\varepsilon _2\mu_2 k^2 - q^2}}
\approx \frac{\mu _2-1}{\mu _2+1}\approx 2\pi n\alpha_H
\]
The friction force acting on a particle moving in parallel to a
plane surface can be obtained as the ratio between the change of
the frictional shear stress between two surfaces after
displacement of body \textbf{2} by small distance $dz$, and the
number of the particles in a slab with thickness $dz$:
\[
F_x^{part}=\frac{d\sigma _{\Vert }(z)}{ndz}\Big |_{z=d}=
\frac {2\hbar} {\pi ^2}\int_0^\infty d\omega \int d^2q\frac{q_xq}{q^2 - \beta^2 q_y^2}e^{-2qd}\left( n_2(\omega^{\prime})-n_1(\omega
)\right)
\]
\begin{equation}
\times\left[q^2(\mathrm{Im}R_p \mathrm{Im}\alpha_E^{\prime} +  \mathrm{Im}R_s \mathrm{Im}\alpha_H^{\prime})
+\beta^2 q_y^2(\mathrm{Im}R_p \mathrm{Im}\alpha_H^{\prime} +  \mathrm{Im}R_s \mathrm{Im}\alpha_E^{\prime})\right]
\label{frparticle}
\end{equation}
where $\alpha_{E(H)}^{\prime}= \alpha_{E(H)}(\omega^{\prime})$.
   For a spherical particle with radius $R$ the electric and magnetic susceptibilities
are given by \cite{LandauEl}
\begin{equation}
\alpha_E = R^3 \frac{\varepsilon -1}{\varepsilon + 2}
\label{elpol}
\end{equation}
\begin{equation}
\alpha_H = -\frac{R^3}{2}\left[1 -\frac{3}{R^2\kappa^2} +
\frac{3}{R\kappa}\cot{R\kappa}\right] \label{mpol}
\end{equation}
where $\kappa = k\sqrt{\varepsilon}$.
Fig. 3 shows the velocity-dependence of the friction force which acts on a small copper particle with $R=10$nm, moving above copper sample at $d=20$nm. The
contributions from the electric dipole and magnetic moments are shown separately. At small velocities the contribution from the magnetic moment is seven
orders of magnitude larger then the contribution from the electric dipole moment.
\begin{figure}
\includegraphics[width=0.45\textwidth]{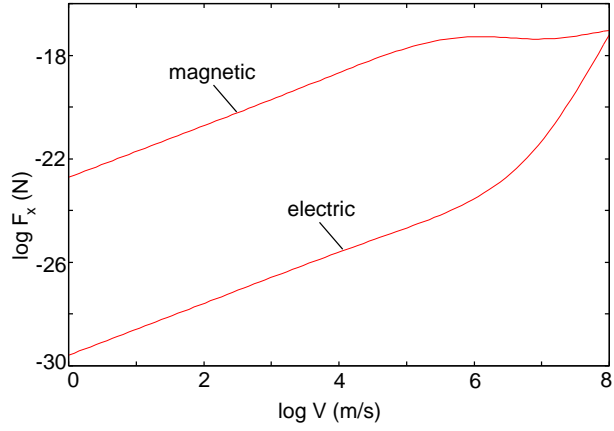}
\caption{\label{Fig3}
The velocity dependence of the friction force acting between small copper particle with radius $R=10$nm moving above a
 copper sample at the separation $d=20$nm. The contributions from the electric dipole moment and the magnetic moment are shown separately.
(The base of the logarithm is 10)
 }
\end{figure}

\section{Calculation of the conservative force between moving bodies \label{force}}

The $z$-component of the force is determined by $zz$-component of the Maxwell stress tensor:
\[
\sigma_{zz} =-\frac 1{4\pi }\int_0^{\infty }d\omega \int \frac{d^2q}{(2\pi)^2}\frac{k_z}{k^2} \left[
(k_z+k_z^{*})(\left\langle \mid w_p\mid
^2\right\rangle +\left\langle \mid w_s\mid
^2\right\rangle
 \right.
\]
\begin{equation}
\left.+\left\langle \mid v_p\mid ^2\right\rangle +
\left\langle \mid v_s\mid ^2\right\rangle) +(k_z-k_z^{*})\left\langle (w_pv_p^* + w_sv_s^*
 + c.c\right\rangle \right]
\label{2four}
\end{equation}
One has to subtract the infinite vacuum contribution to the force which does not depend on separation
$d$ \cite{Lifshitz54,Schwinger78}.
Substituting (\ref{thirteen}-\ref{sixteen}) into  Eq. (\ref{2four}) , and  averaging over the fluctuating electromagnetic field with the help of
Eq. (\ref{seventeen}), and after subtraction  of the vacuum term,  we get the $z$-component of the force:
\[
F_z = \sigma_{zz} =-\frac \hbar {4\pi ^3}\mathrm{Re}\int_0^\infty
d\omega \int d^2q\frac{k_z}{\Delta}e^{2ik_zd} \Big\{(q^2 -\beta
kq_x)^2[R_{1p}R_{2p}^{\prime}D_{ss}
\]
\[
+ R_{1s}R_{2s}^{\prime}D_{pp}] -
\beta^2k_z^2q_y^2[R_{1p}R_{2s}^{\prime}D_{sp} +
R_{1s}R_{2p}^{\prime}D_{ps}]\Big\} [1 + n_1(\omega) +
n_2(\omega^{\prime})]
\]
\[
-\frac \hbar {16\pi ^3}\int_0^\infty d\omega \int_{q<\omega
/c}d^2q\frac{k_z}{|\Delta|^2}[(q^2 - \beta kq_x)^2 +
\beta^2k_z^2q_y^2]
\]
\[
\times  \{(q^2 - \beta kq_x)^2[(1-\mid R_{1p}\mid ^2)(1+\mid
R_{2p}^{\prime }\mid ^2)|D_{ss}|^2-(1\leftrightarrow 2,\omega
\leftrightarrow \omega^{\prime})]
\]
\[
+\beta^2k_z^2q_y^2[(1-\mid R_{1p}\mid ^2)(1+\mid R_{2s}^{\prime
}\mid ^2)|D_{sp}|^2 -(1\leftrightarrow 2,\omega \leftrightarrow
\omega^{\prime})]+ (p\leftrightarrow s)\} \left( n_1(\omega )-
n_2(\omega^{\prime})\right)
\]
\[
+\frac \hbar {4\pi ^3}\int_0^\infty d\omega \int_{q>\omega
/c}d^2q\frac{|k_z|}{|\Delta|^2}[(q^2 - \beta kq_x)^2 +
\beta^2k_z^2q_y^2] e^{-2\mid k_z\mid d}
\]
\[
\times \{(q^2 - \beta kq_x)^2[
\mathrm{Im}R_{1p}\mathrm{Re}R_{2p}^{\prime}|D_{ss}|^2-
(1\leftrightarrow 2,\omega \leftrightarrow \omega^{\prime})]-
 \beta^2k_z^2q_y^2[
\mathrm{Im}R_{1p}\mathrm{Re}
R_{2s}^{\prime}|D_{sp}|^2-(1\leftrightarrow 2,\omega
\leftrightarrow \omega^{\prime})]
\]
\begin{equation}
+ (p\leftrightarrow s)\}\left( n_1(\omega
)-n_2(\omega^{\prime})\right). \label{2five}
\end{equation}

At $T_1=T_2=0$ K, Eq. (\ref{2five}) takes the form
\[
F_z =-\frac \hbar {4\pi ^3}\mathrm{Re}\left\{\int_0^\infty d\omega
\int d^2q - \int_{-\infty}^{\infty}dq_y\int_{0}^{\infty}dq_x
\int_0^{q_xV} d\omega\right\}\frac{k_z}{\Delta}e^{2ik_zd}
\]
\[
\Big\{(q^2 -\beta kq_x)^2[R_{1p}R_{2p}^{\prime}D_{ss}+
R_{1s}R_{2s}^{\prime}D_{pp}]
\]
\[
- \beta^2k_z^2q_y^2[R_{1p}R_{2s}^{\prime}D_{sp} +
 R_{1s}R_{2p}^{\prime}D_{ps}]\Big\}
 \]
  \[
 +\frac \hbar {4\pi ^3}\int_{-\infty}^{\infty}dq_y\int_{0}^{\infty}dq_x
\int_0^{q_xV} d\omega \frac{|k_z|}{|\Delta|^2}[(q^2 - \beta
kq_x)^2 + \beta^2k_z^2q_y^2] e^{-2\mid k_z\mid d}
\]
\[
\times \{(q^2 - \beta kq_x)^2[
\mathrm{Im}R_{1p}\mathrm{Re}R_{2p}^{\prime}|D_{ss}|^2-
(1\leftrightarrow 2,\omega \leftrightarrow \omega^{\prime})]-
 \beta^2k_z^2q_y^2[
\mathrm{Im}R_{1p}\mathrm{Re}
R_{2s}^{\prime}|D_{sp}|^2-(1\leftrightarrow 2,\omega
\leftrightarrow \omega^{\prime})]
\]
\begin{equation}
+ (p\leftrightarrow s)\}\left( n_1(\omega
)-n_2(\omega^{\prime})\right). \label{2six}
\end{equation}
If in Eq. (\ref{2five})  one neglects the terms of the order
$\beta^2$, then, as in the case of friction, the contributions
from the waves with $p$- and $s$- polarization will be separated.
In this case Eq. (\ref{2five})  reduces to
\[
F_z = -\frac \hbar {4\pi ^3}\mathrm{Re}\int_0^\infty d\omega \int
d^2qk_z \left(\frac{1}{R_{1p}^{-1}R_{2p}^{\prime -1}e^{-2ik_zd}-1}
\right.
\]
\[
\left. + \frac{1}{R_{1s}^{-1}R_{2s}^{\prime
-1}e^{-2ik_zd}-1}\right)[1 + n_1(\omega) + n_2(\omega^{\prime})]
\]
\[
-\frac \hbar {16\pi ^3}\int_0^\infty d\omega \int_{q<\omega
/c}d^2qk_z   \Big \{ \frac{[(1-\mid R_{1p}\mid ^2)(1+\mid
R_{2p}^{\prime }\mid ^2)-(1\leftrightarrow 2,\omega
\leftrightarrow \omega^{\prime})]}{|D_{pp}|2}
\]
\[
+ (p\leftrightarrow s) \Big\} \left( n_1(\omega )-
n_2(\omega^{\prime})\right)
\]
\[
+\frac \hbar {4\pi ^3}\int_0^\infty d\omega \int_{q>\omega
/c}d^2q|k_z| e^{-2\mid k_z\mid d}
\]
\begin{equation}
\times \left\{\frac {[
\mathrm{Im}R_{1p}\mathrm{Re}R_{2p}^{\prime}- (1\leftrightarrow
2,\omega \leftrightarrow \omega^{\prime})]} {|D_{pp}|^2} +
(p\leftrightarrow s)\right\}\left( n_1(\omega
)-n_2(\omega^{\prime})\right)
  \label{2seven}
\end{equation}
If  we put $V=0$ and use the Fresnel's formulas for the reflection
amplitudes, then Eq. (\ref{2seven})   reduces to the formula
obtained by Lifshitz \cite{Lifshitz54}. Lifshitz have shown that
at $T_1=T_2=0$ K  it is convenient to transform
$\omega$-integration along the real axis into the integral along
the imaginary axis in the upper half of the complex
$\omega$-plane. For a rarefied body, similarly as in Sec.
\ref{friction}, for $d\ll \hbar/k_BT$  from (\ref{2five}) we get
the van der Waals interaction between a small particle and plane
surface:
\[
F_z = \frac \hbar {\pi ^2}\mathrm{Im}\int_0^\infty d\omega \int
d^2q\frac{q^2e^{-2qd}} {q^2 - \beta^2q_y^2} \Big\{q^2
[R_{p}\alpha_{E}^{\prime}
\]
\[
+ R_{s}\alpha_{H}^{\prime}] + \beta^2q_y^2[R_{p}\alpha_H^{\prime}
+ R_{s}\alpha_E^{\prime}]\Big\} [1 + n_1(\omega) +
n_2(\omega^{\prime})]
\]
\[
+\frac \hbar {\pi ^2}\int_0^\infty d\omega \int d^2q\frac{q^2}{q^2
- \beta^2q_y^2} e^{-2qd}
\]
\[
\times \{q^2 [ \mathrm{Im}R_{p}\mathrm{Re}\alpha_E^{\prime}-
\mathrm{Re}R_{p}\mathrm{Im}\alpha_E^{\prime}]+
 \beta^2q_y^2[
\mathrm{Im}R_{p}\mathrm{Re}
\alpha_H^{\prime}-\mathrm{Im}R_{s}\mathrm{Re} \alpha_E^{\prime}]
\]
\begin{equation}
+ (p\leftrightarrow s,E\leftrightarrow H)\}\left( n_1(\omega
)-n_2(\omega^{\prime})\right).\label{partforce}
\end{equation}

\section{Calculation of the radiative heat transfer between moving bodies \label{heat}}

The radiative energy transfer between the bodies is determined by
 the ensemble average of the Poynting's vector. In the case of  two plane parallel surfaces
the heat flux across the surface \textbf{1} is given by \cite{RMP07}
\[
\left\langle \mathbf{S}_{1z}(\mathbf{r})\right\rangle _\omega =(c/8\pi
)\left\langle \mathbf{E}(\mathbf{r})\times \mathbf{B}^{*}(\mathbf{r}
)\right\rangle _\omega +c.c.
\]
\begin{equation}
=\frac{ic^2}{8\pi \omega }\left\{ \langle \mathbf{E
}(\mathbf{r})\cdot \frac{d}{dz}\mathbf{E}^{*}(\mathbf{r})\rangle
-c.c\right\} _{z=0}. \label{poynting}
\end{equation}
Using Eqs. (\ref{one},\ref{two}) and (\ref{poynting}) we get equations for the heat obtained by the body \textbf{1}
which are very similar
to Eq. (\ref{2two}) and (\ref{2three}):
\[
S_1 =\frac 1{4\pi }\int_0^{\infty }d\omega \int \frac{d^2q}{(2\pi)^2}\frac{\omega}{k^2} \left[
(k_z+k_z^{*})(\left\langle \mid w_p\mid
^2\right\rangle +\left\langle \mid w_s\mid
^2\right\rangle \right.
\]
\begin{equation}
\left. -\left\langle \mid v_p\mid ^2\right\rangle -
\left\langle \mid v_s\mid ^2\right\rangle) +(k_z-k_z^{*})\left\langle (w_pv_p^* + w_sv_s^*
 - c.c\right\rangle \right].
\label{heat1}
\end{equation}
After averaging of the product of the components of the fluctuating electromagnetic field in the same way as in
Sec. (\ref{friction}) we get

\[
S_1  =\frac {\hbar} {8\pi ^3}\int_0^\infty d\omega \int_{q<\omega /c}d^2q\frac{\omega}{|\Delta|^2}[(q^2 - \beta kq_x)^2 + \beta^2k_z^2q_y^2]
\]
\[
\times  [(q^2 - \beta kq_x)^2(1-\mid R_{1p}\mid ^2)(1-\mid R_{2p}^{\prime }\mid ^2)|D_{ss}|^2
\]
\[
+\beta^2k_z^2q_y^2(1-\mid R_{1p}\mid ^2)(1-\mid R_{2s}^{\prime }\mid ^2)|D_{sp}|^2 + (p\leftrightarrow s)]
\left( n_2(\omega^{\prime})-n_1(\omega )\right)
\]
\[
+\frac \hbar {2\pi ^3}\int_0^\infty d\omega \int_{q>\omega
/c}d^2q\frac{\omega}{|\Delta|^2}[(q^2 - \beta kq_x)^2 + \beta^2k_z^2q_y^2]
e^{-2\mid k_z\mid d}
\]
\[
\times [(q^2 - \beta kq_x)^2 \mathrm{Im}R_{1p}\mathrm{Im}R_{2p}^{\prime}|D_{ss}|^2- \beta^2k_z^2q_y^2 \mathrm{Im}R_{1p}\mathrm{Im}
R_{2s}^{\prime}|D_{sp}|^2
\]
\begin{equation}
+ (p\leftrightarrow s)]\left( n_2(\omega^{\prime})-n_1(\omega
)\right).   \label{heat2}
\end{equation}
 Eq. (\ref{heat2}) generalizes the equations for the heat transfer between
two surfaces which are at the rest in the $K$-reference frame \cite{Polder,Volokitin01a}, to the case when the surfaces are moving relative to each
other.
There is also the heat $S_2$ obtained by the body $\textbf{2}$ in the $K^{\prime}$-reference frame. Actually, $S_1$ and $S_2$ are the same quantities,
looked at from different coordinate systems. These quantities are related by the equation:
\begin{equation}
F_xV = S_1 + S_2/\gamma.
\label{heat3}
\end{equation}

For the limiting case of rarefied body, similarly as in Sec.
\ref{friction}, from (\ref{heat2}) we get the contribution from
evanescent waves to the heat adsorbed by a semi-infinite solid at
$d\ll \lambda_T$ in  the $K$-reference frame:
\[
S=
\frac {2\hbar} {\pi ^2}\int_0^\infty d\omega \int d^2q\frac{q\omega}{q^2 - \beta^2 q_y^2}e^{-2qd}\left( n_2(\omega^{\prime})-n_1(\omega
)\right)
\]
\begin{equation}
\times\left[q^2(\mathrm{Im}R_p \mathrm{Im}\alpha_E^{\prime} +  \mathrm{Im}R_s \mathrm{Im}\alpha_H^{\prime})
+\beta^2 q_y^2(\mathrm{Im}R_p \mathrm{Im}\alpha_H^{\prime} +  \mathrm{Im}R_s \mathrm{Im}\alpha_E^{\prime})\right]
\label{heatparticle}
\end{equation}
 The heat adsorbed by a particle can be determined from Eq. (\ref{heat3}).

\section{Calculation of the friction force on a small neutral particle moving relative to black body radiation \label{bb}}

We consider a small neutral particle moving relative to black body
radiation. We introduce two reference frame $K$ and $K^{\prime}$.
The thermal radiation is in equilibrium in the $K$-reference frame
and the particle is at rest in the $K^{\prime}$-reference frames.
We assume that the particle moves with  velocity
 $V$ along the $x$-axis. The relation between the $x$-components of the momentum in the different reference frames is given by
\begin{equation}
p_x=(p_x^{\prime}+\beta E_0/c)\gamma
\label{1bb}
\end{equation}
where $E_0$ is the rest energy of the particle. The rest energy can change due to thermal radiation of the particle. From Eq. (\ref{1bb}) we get
\begin{equation}
\frac{dp_x}{dt}=\frac{dp_x^{\prime}}{dt^{\prime}}+\frac {V}{c^2} \frac{dE_0}{dt^{\prime}}
\label{2bb}
\end{equation}
According to the Einstein law
\begin{equation}
\frac{dm_0}{dt^{\prime}}=\frac {1}{c^2}\frac{dE_0}{dt^{\prime}}
\label{3bb}
\end{equation}
where $m_0$ is the rest mast of the particle. Taking into account that
\begin{equation}
\frac{dp_x}{dt} = \frac{d(m_0V\gamma)}{dt}=\frac{dm_0}{dt^{\prime}}V + m_0\frac{d(V\gamma)}{dt}
\label{4bb}
\end{equation}
from Eqs. (\ref{2bb}-\ref{4bb}) we get
\begin{equation}
m_0\frac{d(V\gamma)}{dt}=\frac{dp_x^{\prime}}{dt^{\prime}}
\label{4bc}
\end{equation}
In the rest reference frame, due to symmetry, the total radiated
momentum  from  the  dipole and magneto-dipole radiation  is
identically zero. Thus, the change of momentum of the particle  in
the rest reference frame is determined by the Lorenz force
$F_x^{\prime}$ acting on the particle from the external
electromagnetic field associated with the  thermal radiation
observed in this reference frame. The dynamics of the particle  in
the $K$-reference frame is determined by the equation
\begin{equation}
m_0\frac{d(V\gamma)}{dt} = F_x^{\prime}
\label{5bb}
\end{equation}
Eq. (\ref{5bb}) does not contain force from the thermal electromagnetic field radiated by the particle.
Thus, from Eq. (\ref{5bb}) it follows
 that, contrary to the claim of the authors of \cite{Dedkov05}, the thermal radiation of the particle
 can not produce any
acceleration. In the $K^{\prime}$-reference frame the Lorenz force
on the particle is determined by the expression
\cite{Volokitin02,Dedkov08}
\begin{equation}
F_x^{\prime} = \frac{\partial}{\partial x ^{\prime}}\langle \mathbf{p_e^{\prime}\cdot E^{\prime *}
(r^{\prime})}\rangle_{\mathbf{r^{\prime}
=r^{\prime}_0}} + \frac{\partial}{\partial x ^{\prime}}\langle \mathbf{p_m^{\prime}\cdot B^{\prime *}(r^{\prime})}
\rangle_{\mathbf{r^{\prime}
=r^{\prime}_0}}
\label{6bb}
\end{equation}
Writing the electromagnetic field as a Fourier integral, and
taking into account that
\[
\mathbf{p}_e^{\prime} =
\alpha_E(\omega^{\prime})\mathbf{E}^{\prime}(\mathbf{r^{\prime}_0})e^{-i\omega^{\prime}t^{\prime}},
\]
\[
\mathbf{p}_m^{\prime} =
\alpha_H(\omega^{\prime})\mathbf{B}^{\prime}(\mathbf{r^{\prime}_0})e^{-i\omega^{\prime}t^{\prime}},
\]
we get
\[
F_x^{\prime} = -i\int_{\infty}^{\infty} \frac{d\omega^{\prime}}{2\pi}\int \frac{d^3k^{\prime}}{(2\pi)^3}k_x^{\prime}
\left[\alpha_E(\omega^{\prime})\langle \mathbf{E^{\prime }\cdot E^{\prime *}}\rangle_{\omega^{\prime}\mathbf{k}^{\prime}} +
\alpha_H(\omega^{\prime})\langle \mathbf{B^{\prime}\cdot B^{\prime *}}\rangle_{\omega^{\prime}\mathbf{k}^{\prime}}\right]
\]
\begin{equation}
=-i\int_{\infty}^{\infty} \frac{d\omega^{\prime}}{2\pi}\int \frac{d^3k^{\prime}}{(2\pi)^3}k_x^{\prime}
\left[\alpha_E(\omega^{\prime}) + \alpha_H(\omega^{\prime})\right]\langle \mathbf{E^{\prime}\cdot E^{\prime *}}\rangle_{\omega^{\prime}\mathbf{k}^{\prime}}
\label{7bb}
\end{equation}
where we have taken into account that for plane waves $\langle
\mathbf{E^{\prime}\cdot E^{\prime *}}
\rangle_{\omega^{\prime}\mathbf{k}^{\prime}}= \langle
\mathbf{B^{\prime}\cdot B^{\prime
*}}\rangle_{\omega^{\prime}\mathbf{k}^{\prime}}$. When we change
from the $K^{\prime}$-reference frame to the
 $K$-reference frame $\langle \mathbf{E^{\prime}\cdot E^{\prime *}}\rangle_{\omega^{\prime}\mathbf{k}^{\prime}}$ is transformed as the energy density of
a plane electromagnetic field. From the law of transformation of the energy density of
a plane electromagnetic field \cite{LandauField} we get
\begin{equation}
\langle \mathbf{E^{\prime}\cdot E^{\prime *}}\rangle_{\omega^{\prime}\mathbf{k}^{\prime}} =
\langle \mathbf{E\cdot E^{ *}}\rangle_{\omega \mathbf{k}}\left(\frac{\omega^{\prime}}{\omega}\right)^2.
\label{8bb}
\end{equation}
According to the theory of the fluctuating electromagnetic field \cite{Lifshitz80}
\begin{equation}
\langle \mathbf{E\cdot E^{*}}\rangle_{\omega \mathbf{k}}
= 4\pi^2 \hbar k \left\{\delta (\frac{\omega}{c}-k) - \delta (\frac{\omega}{c}+k)\right\}[1 + 2n(\omega)]
\label{9bb}
\end{equation}
Taking into account the invariance of the square of the four-wave vector  $(\omega/c)^2 - k^2  = (\omega^{\prime}/c)^2 - k^{\prime 2}$ Eq. (\ref{9bb})
can be rewritten in the form
\begin{equation}
\langle \mathbf{E^{ }\cdot E^{ *}}\rangle_{\omega \mathbf{k}}
= \frac{4\pi^2 \hbar k^2}{k^{\prime}} \left\{\delta (\frac{\omega^{\prime}}{c}-k^{\prime}) - \delta (\frac{\omega^{\prime}}{c}+k^{\prime})\right\}
[1 + 2n(\omega)]
\label{10bb}
\end{equation}
Substitution Eqs. (\ref{8bb},\ref{10bb}) in Eq. (\ref{7bb}) and integration over $\omega^{\prime}$
gives
\[
F_x = \frac{\hbar c}{2\pi^2}\int d^3k k k_x \left[\mathrm{Im}\alpha_E (ck) + \mathrm{Im}\alpha_H (ck)\right]
\]
\begin{equation}
\times \left\{n[\gamma(ck+Vk_x)] -
n[\gamma(ck-Vk_x)]\right\},
\label{12bb}
\end{equation}
where we have omitted index of prime and have taken into account that $\omega = (\omega^{\prime} +
k^{\prime}_xV)\gamma$.
Introducing the new variable $\omega = ck$,  (\ref{12bb}) can be transformed to the form
\[
F_x = \frac{2\hbar }{\pi c^2}\int_0^{\infty} d\omega \omega^2 \int_0^{\omega/c} dk_x k_x
\left[\mathrm{Im}\alpha_E (\omega) +
\mathrm{Im}\alpha_H (\omega)\right]
\]
\begin{equation}
\times \left\{n[\gamma(ck+Vk_x)] -
n[\gamma(ck-Vk_x)]\right\}
\label{13bb}
\end{equation}
Eq. (\ref{13bb}) generalizes the result obtained in
\cite{Mkrtchian} to the case of large velocities, and includes the
contribution from the magnetic moment. For metallic particles the
contribution from the magnetic moment exceeds substantially the
contribution from the electric dipole moment.   At small
velocities $F_x = -\Gamma V$, where
\begin{equation}
\Gamma = \frac {4\hbar}{3\pi c^5}\int_0^{\infty}d\omega \left(-\frac{\partial n}{\partial
\omega}\right)\omega^5[\mathrm{Im}\alpha_E(\omega) +
\mathrm{Im}\alpha_H(\omega)]
\label{14bb}
\end{equation}
For metals with $4\pi\sigma \gg k_BT/\hbar$ and for $c\sqrt{2\pi \sigma k_BT}\gg R$, where $\sigma $ is the
conductivity, from Eqs. (\ref{elpol}) and
 (\ref{mpol}) we get
\begin{equation}
\mathrm{Im}\alpha_E(\omega) \approx R^3\frac {3\omega}{4\pi \sigma}
\label{15bb}
\end{equation}
\begin{equation}
\mathrm{Im}\alpha_H(\omega) = \frac {4\pi \sigma \omega R^5}{30c^2}
\label{16bb}
\end{equation}
Setting the friction coefficient $\Gamma$ to $m_0/\tau$, where $\tau$ the relaxation  time, and using
$m_0 = 4\pi R^3\rho/3$, from Eqs. (\ref{14bb}
-\ref{16bb}) we get
\begin{equation}
\tau^{-1}_e \approx 10^2\frac {\hbar}{\rho \lambda_T^5}\frac {k_BT}{\hbar \sigma}
\label{17bb}
\end{equation}
\begin{equation}
\tau^{-1}_m \approx 10^2\frac {\hbar R}{\rho \lambda_T^6}\frac {\sigma R}{c}
\label{18bb}
\end{equation}
where  $\tau_e^{-1}$ and $\tau_m^{-1}$ are the contributions to
the friction from the electric dipole and  magnetic moments,
respectively. For $T= 300$ K, $\rho \approx 10^4$ kg/m$^3$,
$\sigma \approx 10^{18}$ s$^{-1}$ from Eqs.
(\ref{15bb},\ref{18bb}) we get $\tau_e \sim 10^{16}$ s and $\tau_m
\sim 10^{12}$ s. When the conductivity decreases $\tau_e$ also
decreases and reaches minimum at $2\pi \sigma \approx k_BT/\hbar$.
At $T=3000$ K this minimum corresponds to about a day
($\tau_e^{min} \approx 10^5$ s). In \cite{Mkrtchian} the same
relaxation time was obtained for $Ba^+$.

\section{Conclusion \label{conclusions}}

In this paper  within the framework of unified approach we have
calculated the Casimir-Lifshitz interaction,  the van der Waals
friction force and the radiative heat transfer at nonequilibrium
conditions, when the interacting bodies are at different
temperatures, and they move relative to each other with the
arbitrary velocity $V$. In comparison with the existing literature
we have studied more general nonequilibrium conditions. Our study
was focused on the surface-surface and surface-particle
configuration. We have found the exact solution of problem about
the determination of the fluctuating electromagnetic field in the
vacuum gap between two flat parallel surfaces moving relative to
each other. Knowing the electromagnetic field we have calculated
the Maxwell stress tensor and the Poynting vector which determine
the friction and conservative forces, and the heat transfer
between the solids, respectively. For the heat transfer and the
conservative force our treatment generalizes the results obtained
for bodies at rest to the case of bodies which move relative to
each other. We have  found that the velocity dependence of the
considered phenomena  is determined by Doppler shift and can be
strong for resonant photon tunneling between the surface modes.
This effect can be used for the precision determination of energy
of the surface modes. We have shown that relativistic effects
produce a mixing of the $s$ and $p$-wave contributions to the
forces and the heat transfer. This relativistic effect is of the
order $(V/c)^2$. If one neglects by terms of  order $(V/c)^2$, the
different polarizations will contribute to the forces and the heat
transfer separately. The velocity dependence of the conservative
force is much weaker than the velocity dependence of the friction
force and the heat transfer.  For the conservative force the
important range of the frequencies is determined by the plasma
frequency $\omega_p$ which is much larger than the Doppler shift
$\Delta \omega =q_xV\sim V/d$ for practically  all separations and
velocities. Therefore the velocity dependent  component of the
Casimir-Lifshitz force  will be considerably smaller  than the
$V=0$ component.  The same is true for the thermal component of
the Casimir-Lifshitz force. The thermal component was measured
recently in \cite{Antezza05}. Thus, one can expect that the
velocity dependent component also can be measured. For the
friction force and the heat transfer the important range of the
frequencies is determined by the thermal frequency
$\omega_T=k_BT/\hbar$ which can be of the same order as the
Doppler shift. The friction force initially increases when the
velocity increases, reaches maximum at $V\sim \omega_T$, and then
decreases. From the limit when one of the bodies is rarefied,  we
have calculated the interaction force
 and the heat transfer for a small particle outside a flat surface. For a particle we have taken into account
the contribution to the forces and the heat transfer from the
electric dipole and magnetic moments. For metallic particles we
have found that the contribution to the friction from the magnetic
moment exceeds the contribution from the electric dipole moment by
several  orders of  magnitude. We have presented a relativistic
theory of the friction force acting on a particle moving relative
to the black body radiation. We have shown that the thermal
radiation of the particle does not produce any acceleration and
that the friction force acting on the particle is determined by
external electromagnetic field associated with  the black body
radiation. For a  metallic particle moving relative to the black
body radiation the friction force increases initially when the
conductivity decreases,  reaches maximum at $2\pi\sigma \approx
\omega_T$, and then decreases.

\vskip 0.5cm

A.I.V acknowledges financial support from the Russian Foundation
for Basic Research (Grant N 08-02-00141-a) and  DFG.


\vskip 0.5cm







\begin{thebibliography}{999}


\bibitem{Rytov53}  S. M. Rytov, \textit{Theory of Electrical Fluctuation and
Thermal Radiation} (Academy of Science of USSR Publishing, Moscow, 1953)

\bibitem{Rytov67}  M. L. Levin, S. M. Rytov, \textit{Theory of eqilibrium thermal
fluctuations in electrodynamics} (Science Publishing, Moscow, 1967)

\bibitem{Rytov89}  S. M. Rytov, Yu. A. Kravtsov, and V. I. Tatarskii, \textit{
Principles of Statistical Radiophyics}(Springer, New York.1989), Vol.3




\bibitem{Lifshitz54}  E. M. Lifshitz, Zh. Eksp. Teor. Fiz. \textbf{29 }94
(1955) [Sov. Phys.-JETP \textbf{2 }73 (1956)]

\bibitem{Polder}  D. Polder and M. Van Hove, Phys. Rev. B \textbf{4}, 3303
(1971)



\bibitem{Volokitin99}  A. I. Volokitin and B. N. J. Persson, J.Phys.:
Condens. Matter \textbf{11}, 345
(1999);Phys.Low-Dim.Struct.\textbf{7/8},17 (1998)


\bibitem{Pitaevskii08} M. Antezza, L. P. Pitaevskii, S. Stringari,
and V. B. Svetovoy, Phys. Rev. A \textbf{77}, 022901 (2008)

\bibitem{Antezza05} M. Antezza, L. P. Pitaevskii, and S. Stringari,
Phys. Rev. Lett.  \textbf{95}, 113202 (2005)

\bibitem{Antezza06} M. Antezza, L. P. Pitaevskii, S. Stringari, and V. B. Svetovoy,
and Phys. Rev. Lett.  \textbf{97}, 223203 (2006)






\bibitem{Pendry99}  J. B. Pendry, J.Phys.:Condens.Matter \textbf{11}, 6621
(1999).

\bibitem{Volokitin01a}  A. I. Volokitin and B. N. J. Persson, Phys. Rev. B \textbf{
63}, 205404 (2001); Phys. Low-Dim. Struct. \textbf{5/6}, 151 (2001)

\bibitem{Volokitin04}  A. I. Volokitin and B. N. J. Persson, Phys. Rev. B \textbf{%
69}, 045417 (2004)

\bibitem{Volokitin03}  A. I. Volokitin and B. N. J.Persson, JETP Lett. \textbf{78}
, 457(2003)

\bibitem{Mulet01}  J. P. Mulet, K. Joulin, R. Carminati, and J. J. Greffet, Appl.
Phys.Lett. \textbf{78}, 2931 (2001)

\bibitem{Mulet02}  J. P. Mulet, K. Joulain, R. Carminati, and J. J. Greffet, Microscale
Thermophysical Engineering, \textbf{6}(3), 209 (2002)

\bibitem{Kittel}  A. Kittel, W. M\"uller-Hirsch, J. Parisi, S. Biehs,
D. Reddig, and M. Holthaus, Phys. Rev. Lett. \textbf{95}, 224301
(2005).


\bibitem{Joulain05}  K. Joulain, J. P. Mulet, F. Marquier, R. Carminati, and
J. J. Greffet, Surf. Sci. Rep. , \textbf{57, }59 (2005)





\bibitem{RMP07} A. I. Volokitin and B. N. J. Persson, Rev. Mod. Phys. \textbf{79}, 1291 (2007)


\bibitem{Volokitin01b}  A. I. Volokitin and B. N. J. Persson, J.Phys.:
Condens. Matter \textbf{13}, 859 (2001)

\bibitem{Volokitin03a}  A. I. Volokitin and B. N. J. Persson, Phys. Rev. Lett.
\textbf{91}, 106101 (2003).

\bibitem{Volokitin03b}  A. I. Volokitin and B. N. J. Persson, Phys. Rev. B,
\textbf{68}, 155420 (2003).




\bibitem{Mkrtchian} V. Mkrtchian, V. A. Parsegian, R. Podgornik, and W. M. Saslow, Phys. Rev. Lett.
\textbf{91}, 220801 (2003)


\bibitem{Chapius08} P.-O. Chapius, M. Laroche, S. Volz, and J.-J. Greffet, Phys. Rev. \textbf{77}, 125402
(2008)



\bibitem{Pendry97}  J. B. Pendry, J. Phys. C\textbf{9} 10301 (1997).



\bibitem{LandauEl} L. D. Landau and E. M. Lifshitz, \textit{Electrodynamics of
Continuous Media}, Pergamon, Oxford, 1960.




\bibitem{Schwinger78}  J. Schwinger, L. DeRaad, K. Milton, Ann. Phys. \textbf{115}, 1 (1978)




\bibitem{Dedkov05} G. V. Dedkov, A. A. Kyasov, Phys. Lett. A \textbf{339}, 212 (2005)

\bibitem{Volokitin02}  A. I. Volokitin and B. N. J. Persson, Phys. Rev. B \textbf{
65}, 115419 (2002)

\bibitem{Dedkov08} G. V. Dedkov, A. A. Kyasov, Tech. Phys. \textbf{53}, 389 (2008)

\bibitem{LandauField} L. D. Landau and E. M. Lifshitz, \textit{The classical theory of field},
Pergamon, Oxford, 1975.











\bibitem{Lifshitz80} E. M. Lifshitz and L. P. Pitaevskii, \textit{Statistical Physics, Pt.2}
Pergamon, Oxford, 1980.


\end{thebibliography}
\end{document}